# Evolution of magnetic dynamics in an artificially frustrated Fe nanoparticle system

Satyendra Prakash Pal[1,2]* & P. Sen[1]

[1]School of Physical Sciences, Jawaharlal Nehru University, New Delhi-110067, India

[2]Department of Physical Sciences, Indian Institute of Science Education and Research, Knowledge city, Sector 81, SAS Nagar, Manauli-140306, Punjab, India

*email: sppal85@gmail.com

**Frustrated lattices[1-3], characterized by minor breakdown in local order in an otherwise periodic lattice, lead to simultaneous possibilities of several ground states which can trigger unique physical properties, in condensed matter systems. In magnetic materials with atomic spins, frustration takes another shape with added possibilities to construct various topological arrangements of spins, whereby magnetic order is disturbed[2,4]. We have achieved a new approach to introduce positional atomic disorder inside a Fe nanoparticle lattice, forming domains without boundaries to study magnetic dynamics of the constituent spins. This magnetism overrides the exchange bias derived magnetic enhancement, appears only at temperatures around 200 K and is characterized by a dynamic polarity, p = ±1 (positive or negative) with a precise frequency. The material otherwise behaves like a superparamagnet with characteristic magnetization behaviour at room temperature and 2 K.**

Materials such as polymers, cholesteric liquid crystal systems, glassy and amorphous materials, are unique on account of their ability to co-exist in periodic and defected forms[1,2], which designate these as frustrated. There is intense recent activity to understand and demonstrate evolution of magnetic frustration, through unusual synthesis, artificial reproduction of lattices with trapped ions or atoms and synthesis with coupled lasers, in a methodology akin to a bottom-up approach, to simulate a near-ideal situation [5-8]. However, geometric frustration arises in real materials when the topology of a magnetic lattice is disturbed and can accommodate the criteria- frustration and modified positional order (randomness)[3]. These materials can become promoters of new and exotic behaviour.

A few years ago[9], some of us investigated the dynamics of electro-explosion of wires (EEW) employing a 'single wire single explosion' event, whereby a metal needle is exploded against a metal plate, through passage of high current densities. This experiment correlated the energy deposition efficiency to a metallic lattice, in the EEW process, to known material properties such as electron-phonon coupling. Thus in processes where electrons initially achieve large kinetic energies (EEW, ion irradiation etc.), the energy is dissipated primarily to the lattice, in materials with large electron-phonon (e-ph) coupling but low electronic and thermal conductivities (e.g., for Fe, $\lambda_{el-ph}$ = 49.8×10$^{11}$ W cm$^{-3}$K$^{-1}$, $\sigma_{Fe}$ = 102.87 m$\Omega^{-1}$ cm$^{-1}$, and $\kappa_{Fe}$ = 72.8 J/m s deg) and to other lattice electrons, through electron-electron (e-e) scattering in materials with low electron-phonon coupling but high electronic and thermal conductivities (e.g., for Cu, $\lambda_{el-ph}$ = 4.94×10$^{11}$ W cm$^{-3}$K$^{-1}$, $\sigma_{Cu}$ = 595.8 m$\Omega^{-1}$ cm$^{-1}$, and $\kappa_{Fe}$ = 401 J/m s deg). Similar observations have also been made in high energy ion irradiation (hundreds of MeV) experiments, whereby localized disorder can be achieved in an otherwise ordered lattice[10]. The primary driving force for this is nonlinear transport of energy[11] through chains of lattice atoms, which can couple to the lattice when conditions are sufficient.

Here we employ the EEW technique, which allows breakdown of a Fe nanoparticle lattice into smaller ordered units, in what can be called a top-down approach. We show evidence of enhanced magnetism when single-domain nanometer-sized Fe particles are broken down into smaller domains, without interfaces, resulting in correlated dynamic magnetic behaviour. The breakdown in local lattice order which allowed formation of domains, without physical boundaries, was made possible through a nonequilibrium process when the particles were constituted, catalyzed by a strong e-ph coupling in Fe.

The Fe-nano employed (Fig. 1) are almost spherical in shape with a skewed-normal size distribution; most probable size = 7.5 nm and FWHM = 7.5 nm. These particles are thus expected to demonstrate superparamagnetism. The transmission electron microscopy (TEM) data shows disruptions in lattice periodicity (Fig. 1a) and the selected area electron diffraction (SAED) (data not shown) does not return any clear diffraction features. However, an interplanar distance of 0.202 nm, measured at the clear part of the particle, reveal these to be the (110) planes of bulk Fe. The x-ray diffraction (XRD) of the same sample (Fig. 2a) record a prominent feature at $2\theta= 44.8^0$ (the (110) reflection of Fe), minor reflections, together with a large background signal. In general, all peak intensities are fairly low, and with TEM evidence, establish a partial loss in lattice periodicity. The inset allows a feel for the original Fe material, the XRD line shapes and new peaks.

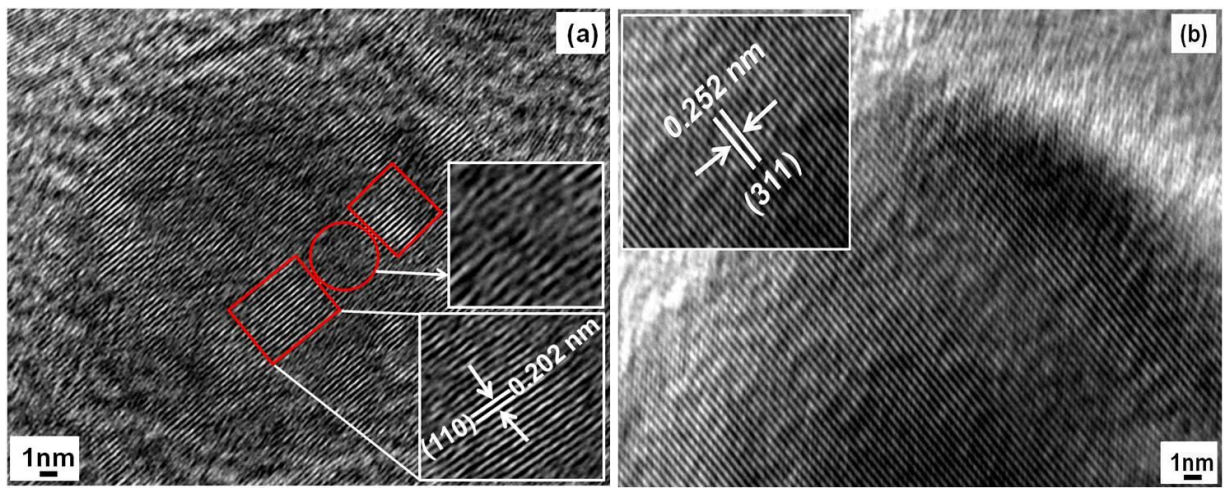

**Figure 1| TEM micrographs of Fe-nano and Fe-nano-ox annealed at 1,000 K**. **a,** TEM image of Fe-nano with periodic lattices (red boxes) shown interspersed with a positional disordered lattice (red circle, enlarged) all in a single nanoparticle, together with similar such regions. Inset records an interplanar distance of 0.202 nm, identified as the (110) planes of pure Fe. **b,** TEM image of Fe-nano-ox annealed at 1,000 K, with enhanced lattice order in the nanoparticle bulk, which extends to a partially ordered outer capping layer. Inset records an interplanar distance of 0.252 nm which correspond to the (311) planes of $Fe_3O_4$ or $\gamma$-$Fe_2O_3$ (see text).

The nanoparticles are fairly resistant to thermal activation as seen in Fig. 2a, following a 1,000 K anneal of Fe-nano. The pure Fe phase is still evident ($2\theta= 44.8^0$) with minor increase in the oxide component, due to rearrangement in the oxide capping layer. Thus, a customary method to repair defected lattices through thermal anneal[12], fail. We adopt an unusual route to infuse lattice periodicity by way of anneal of Fe-nano in the presence of activated carbon (Fe-nano-ox, solid-solid mixture, 1:1 by weight). The XRD of this mixture, following anneal to 1,000 K, clearly demonstrate the success of the intended step. The Fe (110) reflection has been observed to disappear above 600 K, while a major transformation occurs, seen here after anneal at 1,000 K. Accordingly, the various peaks presented in the figure can be assigned to the oxides $\gamma$-$Fe_2O_3$ or $Fe_3O_4$, verified as the former by Mössbauer spectroscopy (see Supplementary Information). Thus following anneal above 600 K, in presence of activated carbon, transformations are induced to crystallize a certain phase or a new crystalline order. TEM of Fe-nano-ox annealed at 1,000 K (Fig. 1b) show distinct atomic arrangements whereby the (311) planes are identified in the figure. Also, the outer capping layer attains clarity, lattice ordering, after anneal (compare Figs. 1a and 1b). Infusion of crystalline order impacts the dynamic magnetic evolution, as reported later.

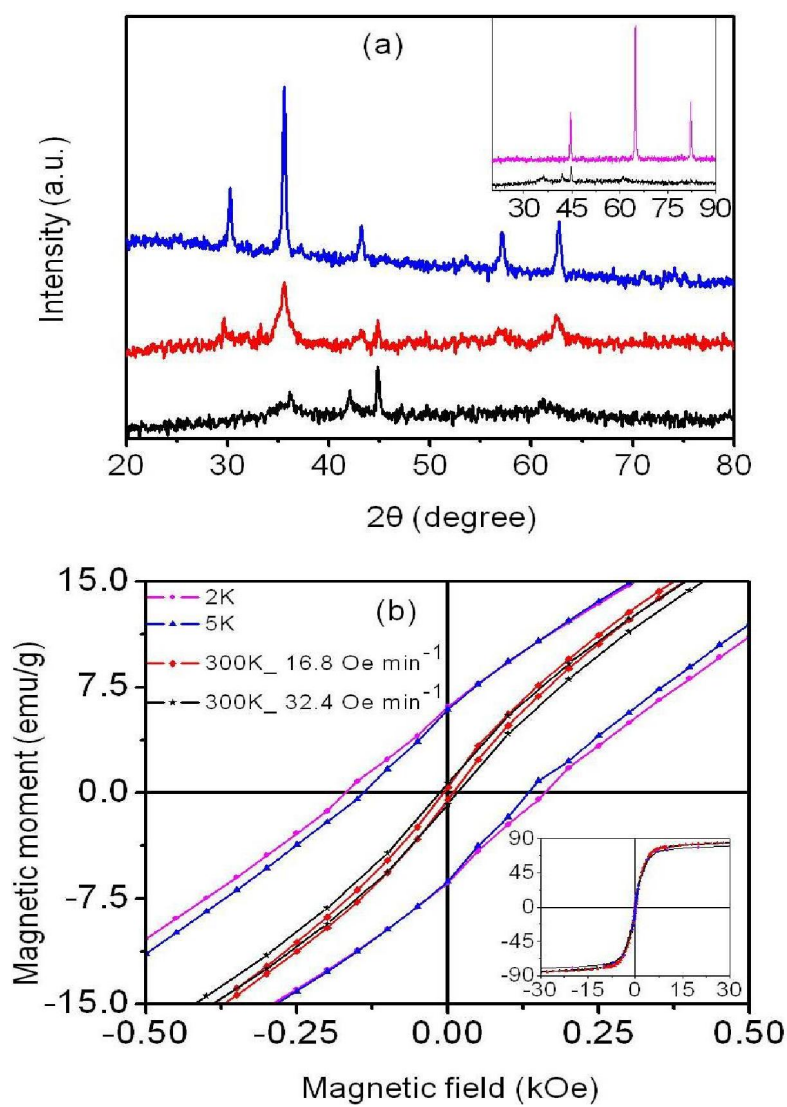

**Figure 2|** **XRD patterns for Fe-nano, annealed Fe-nano and Fe-nano-ox together with hysteresis loops for Fe-nano at various temperatures and magnetic field sweep rates. a,** Lower plot, XRD of Fe-nano with a prominent peak at 2θ = 44.8$^0$, minor reflections at 35.5$^0$ (311 plane) and 62$^0$ (440 plane) due to the capped layer correspond to the cubic system of Fe oxide, γ-Fe$_2$O$_3$ (see text). Middle plot, Fe-nano annealed at 1,000 K, with increased oxide peak intensities. Upper plot, Fe-nano-ox annealed at 1,000 K, with the most intense XRD pattern. Inset compares XRD of the original Fe plate (upper plot) and Fe-nano materials. **b,** M-H loops for Fe nano at 2 K, 5 K and 300 K at a sweep rate of 16.8 Oe min$^{-1}$ is compared with those at 32.4 Oe min$^{-1}$ (300 K). Inset, shows the full M-H curve, swept through ± 30 kOe.

Magnetic hysteresis data (Fig. 2b) provides first information on the magnetic state of Fe-nano. The hysteresis (M-H) curves recorded at 300 K, 5 K and 2 K, are shown for the region around H = 0 when the magnetic field is swept at 16.8 Oe min$^{-1}$. The inset to this figure shows record of the full M-H curve swept through ±30 kOe. This procedure yields 3 coercive field values, 7.8 Oe (300 K), 136.5 Oe (5 K) and 165.4 Oe (2 K) which reflect the slowdown in the dynamics of the magnetic moments, with temperature. However, an enhanced hysteresis is present when data is acquired at a faster sweep rate (32.4 Oe min$^{-1}$, Fig. 2b), at 300 K, returning a coercive field of 15.1 Oe. At an even larger sweep rate (60 Oe min$^{-1}$), at 300 K, this value is 139.0 Oe (see Supplementary Information). These observations indicate the presence of an energy barrier and inherent magnetization dynamics (slow compared to the sweep rate), which slows down with temperature. The following observations are made. The presence of hysteresis, recorded at a faster rate, indicates inherent slow magnetization dynamics and an energy barrier. And, the observed coercivity, which is both temperature- and sweep rate dependent, indicate dynamic behaviour.

The time (t) dependence of the magnetization, M(t), provides spin relaxation dynamics and signify dissipation of stored magnetic energy by constituent assemblies. We measure M(t) at various temperatures, T = 50 K and 200 K, after ramp up of the sample temperature to 330 K, followed by cool down to set values, in the presence of an applied magnetic field (20 kOe). The field is then switched off and M(t) recorded. The data at 50 K show relaxation behaviour typical at these temperatures (Fig. 3a). The experimental points can be fitted to $M(t) = M_0 + A_1 e^{-t/\tau_1} + A_2 e^{-t/\tau_2}$. The decay behaviour is ascribed to a dilute ensemble of superspins with random spatial distribution, anisotropy, and spin sizes[13]. In the lower panel of Fig. 3a, M(t) is plotted to investigate relaxation behaviour, now at an elevated temperature of 200 K. A fit to the above criterion (Fig. 3a, lower panel) is not possible beyond t=3,000 s, as a certain deviation is noticed, which exists for all data points, and consequently, the fit is shown truncated. The fitted form is $M(t)=M_0+A_1e^{-t/\tau_1}$.

An inspection of the deviation reveal an inherent periodicity in variation of M(t), superimposed on the exponential decay which we subject to Fourier analysis (FFT). This procedure allows separation of various frequency components hidden in the compound waveform M(t), to ease detection. This results in the observation of a distinct frequency at $6.60 \times 10^{-4}$ Hz (Fig. 3b). The frequency extracted demonstrates a rather slow periodic variation of magnetization with time, while the enhancement/reduction in magnetic moment, superimposed on the exponential decay, signifies magnetic ordering. Enhancement has indeed been reported before[14] and assigned likewise, however, no oscillations have ever been recorded.

The inherent magnetization dynamics in the decaying part of M(t) in Fig. 3a, the presence of temperature and scan rate dependent hysteresis and an energy barrier, $\Delta E_B$ (Fig. 2b), allows the spin relaxation to be described by $\tau=\tau_0 e^{\Delta E_B/kT}$, which relates the time taken ($\tau$) for magnetization along a defined axis (p = ± 1) to jump in phase (p = ± 1→ p = ± 1), assuming a coherent magnetic reversal. A way to achieve slow down of the relaxation process (large $\tau$) is to lower kT (usual) or increase $\Delta E_B$ (not usual). The latter, however, is achieved in our experiment through the formation of regions within the nanoparticle boundary where lattice periodicity is disturbed (Fig. 1), which can lead to geometric frustration. Competing possibilities introduced, will proliferate ground states and subsequent loss in long range magnetic order. A similar situation has been reported to evolve regions of large magnetic anisotropy[8]. The magnetization overcomes this large anisotropy barrier, slow down as a result, to oscillate at a frequency determined by FFT of the M(t) data. The activation is thermal and appears only above a certain energy (kT) when superspins are rendered free. On the basis of the character of the fit, there are no other exotic spin dynamics present.

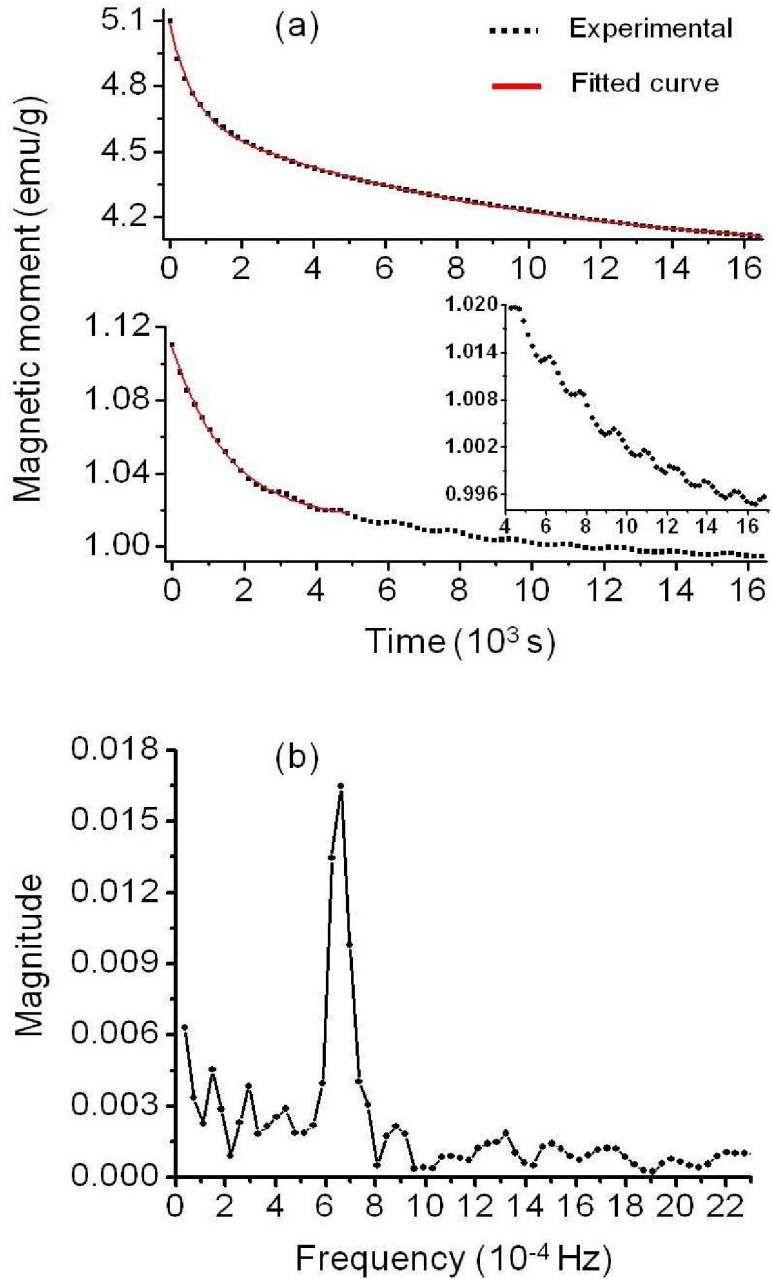

**Figure 3| Magnetic relaxation curves for Fe-nano at 50 K and 200 K and FFT for the 200 K relaxation data. a,** Upper panel, the experimental M(t) data at 50 K, fitted with a combination of two exponentials is shown ($\tau_1$ = 630 s and $\tau_2$ = 9,303 s). Lower panel, shows the M(t) behaviour at 200 K and single exponential fit up to 5,000 s ($\tau_1$ = 1,585 s). The relaxation data at 200 K, shows periodic variation of M(t) superimposed on the magnetic decay (inset). **b,** FFT, derives the frequency associated with M(t).

The contribution of geometric frustration towards a precisely fluctuating magnetic moment is addressed next. The nanoparticles can be ordered through an anneal process at 1,000 K (Fig. 1b), which leads to an oxidized but ordered nanoparticle. The M(t) data for the oxidized nanoparticle, Fe-nano-ox, is shown in Fig. 4 and is immediately recognized from the magnetization M(t=0) of 1.50 emu/g as compared to 1.11 emu/g for Fe-nano at 200 K (Fig. 3). This enhancement can be ascribed to an exchange bias[15], possibly spin frustration[16], following $\gamma$-$Fe_2O_3$ formation. The relaxation can be fitted with an exponential as before. In the inset, we plot the relaxation data after 4,000 s. As before (Fig. 3), the deviation from the exponential fit is clearly evident here, whose magnitude is much reduced on inspection. In order to quantify this deviation, we define a dimensionless quantity $\Delta^{MF} = \frac{\sum |\Delta M|}{(M_{t=initial} + M_{t=final})/2}$, which reflects the impact of geometry derived magnetic frustration ($\Delta^{MF}$) on the deviation ($\Delta M$) from the usual exponential relaxation of magnetic spins. A larger value thus reflects a preponderance of magnetic frustration. For Fe-nano this value is 0.04, while Fe-nano-ox is much reduced at 0.01. This correlates the qualitative differences in lattice order (Figs. 1 and 2) to a recordable quantity through magnetization. The FFT derived from M(t) of Fe-nano-ox is shown in Fig. 4b, records a frequency = $6.23 \times 10^{-4}$ Hz, but considerably weakened. This demonstrates lack of a completely ordered lattice (magnetic).

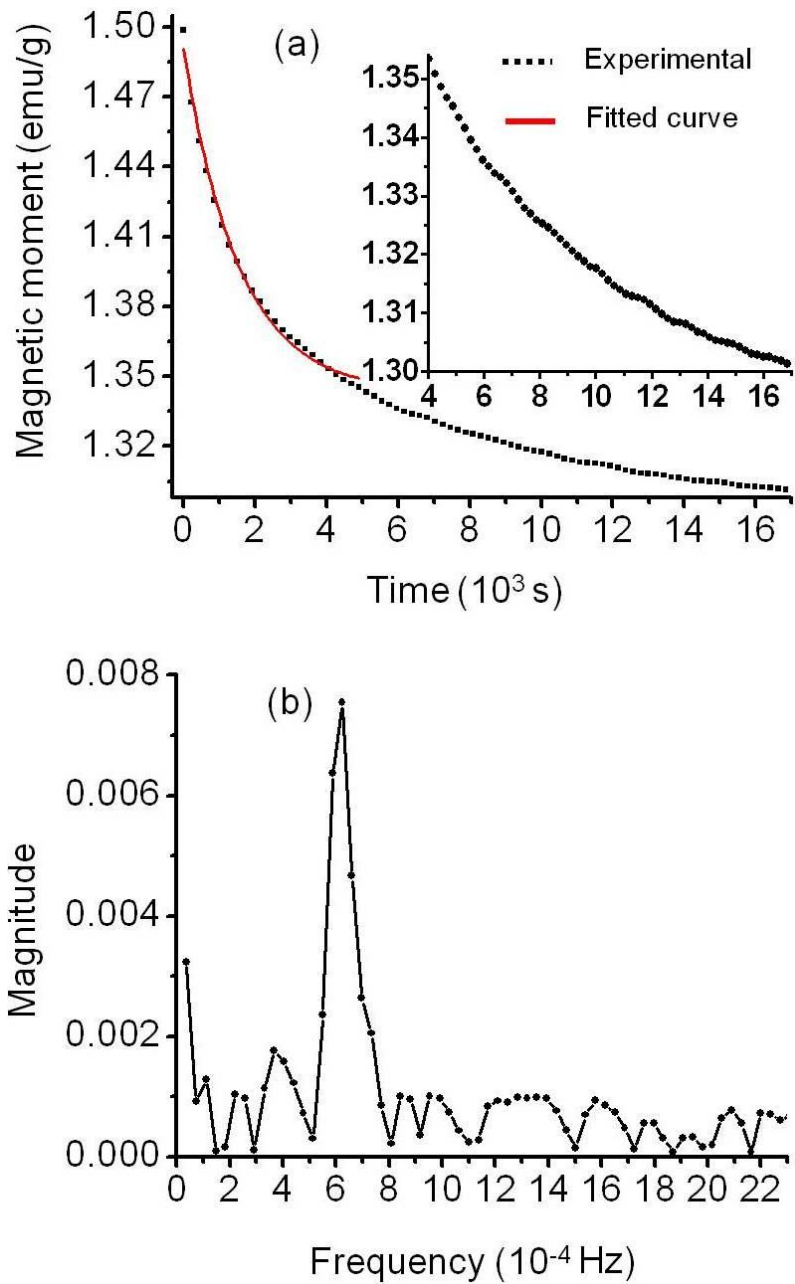

**Figure 4| Magnetic relaxation curves for Fe-nano-ox (1,000 K annealed) at 200 K and FFT for the 200 K relaxation data. a,** The experimental M(t) is shown fitted with a single exponential ($\tau_1$ =1,549 s ), up to 5,000 s. Inset, M(t) after a lag of 4,000 s from t=0. **b,** FFT derived from M(t), demonstrates a periodic variation of M(t) at 200 K.

In summary, we have shown how an inherently nonlinear and magnetically bistable system generate coherent oscillations of its constituent magnetic moments, in the presence of a certain positional disorder in the lattice. This disorder is argued to generate large magnetic anisotropy regions, and demonstrated to be an important step for the oscillations to appear. The M(t) experiments described here present a time evolution map of how a nonlinear system self-organizes. Many such systems in the natural sciences and other fields are known to harbour incipient oscillations called `stable limit cycles'[17]. The coherent oscillations of the magnetic moments recorded in this artificially modified magnetic material can be used to initiate efficient coupling of energy through a resonant processes, to introduce resonant magnetic hyperthermia for treatment of cancer, which can be highly efficient. Moreover, the frequency of the oscillations can be tuned by changing the barrier height $\Delta E_B$, i.e., the magnetic anisotropy energy, through selected EEW parameters. Thus smaller heights will make possible tunable nanosized high frequency oscillators for signal generation and detection, while larger barrier heights will yield a new magnetic order which will change only very slowly, once flipped. An important component of this new magnetic order is its existence under ambient conditions and presence overriding the exchange bias magnetism. This can yield new possibilities for magnetic storage under room temperature conditions for high density data recording.

## METHODS

The material considered for this study are pure nanoparticles of Fe prepared by way of EEW[9, 18]. This method allows control over the purity of the nanoparticles and contaminants in the surface layer, through a simultaneous capping/passivation process while the explosion is carried out in a specified medium, here de-ionized water. The medium, which collects the nanoparticle sample, is finally driven off through evaporative drying, leaving the nanoparticles in powdered form. High purity Fe wires (Sigma Chemicals, product code-266221) of thickness 250 microns were exploded against a Fe plate of similar purity (Sigma Chemicals, product code-356778). Pulsed and d.c. voltages in the range 15–60 V were investigated. The material reported here was obtained at $V_{dc}$ = 40 V and a set current of 50 A, with a variable in-line inductance (20-200 µH) held to cut-off any tailing voltages, following explosion. The wire was driven by a semi-automatic rack and pinion arrangement till a sensing alert is activated on plate contact, when the exploding voltages are triggered with a microcontroller driven solid state switching device. Fe-nano-ox is a composite of Fe nano and activated carbon, 1:1 by weight which is employed to oxidize the nanoparticles and establish lattice order through an annealing step up to 1,000 K. All annealing procedures reported here were carried out in a flow muffle furnace at the temperatures indicated. All temperatures were achieved by ramping up from room temperature with a step size of 5 K min$^{-1}$, in nitrogen atmosphere; once achieved, the samples are held for half an hour at that temperature and cooled to room temperature.

**References:**

1. Sadoc, J.- F. & Mosseri, R. *Geometrical Frustration* (Cambridge University Press, 2006).

2. Gardner, J. S. Geometrically frustrated magnetism. *J. Phys. Cond. Matter.* **23**, 160301 (2011).

3. J. A. Mydosh, *Spin Glasses—An Experimental Introduction* (Taylor and Francis, London, 1993) p. 3.

4. Greedan, J. E., Derakhshan, S., Ramezanipour, F., Siewenie, J., & Proffen Th. A search for disorder in the spin glass double perovskites $Sr_2CaReO_6$ and $Sr_2MgReO_6$ using neutron diffraction and neutron pair distribution function analysis. *J. Phys. Cond. Matter.* **23**, 164213 (2011).

5. Islam, R. *et al.* Emergence and frustration of magnetism with variable-range interactions in a quantum simulator. *Science* **340,** 583-587 (2013).

6. Nixon, M., Ronen, E., Friesem, A. A., & Davidson, N. Observing geometric frustration with thousands of coupled lasers. *Phys. Rev. Lett.* **110**, 184102 (2013).

7. Jo, G.-B. *et al.* Ultracold atoms in a tunable optical kagome lattice. *Phys. Rev. Lett.* **108**, 045305 (2012).

8. Lee, S. H. *et al.* Emergent excitations in a geometrically frustrated magnet. *Nature* **418,** 856-858 (2002).

9. Vandana & Sen, P. Concentric ring patterns in needle–plate exploding system. *J. Phys. Condens. Matter.* **19,** 016009 (2007).

10. Sen, P., Aggarwal, G., & Tiwari, U. Dissipative structure formation in cold-rolled Fe and Ni during heavy ion irradiation. *Phys. Rev. Lett.* **80**, 97-100 (1998).

11. Sen, P., Akhtar, J., & Russell, F.M. MeV ion-induced movement of lattice disorder in single crystalline silicon. *Europhys. Lett.* **51**, 401-405 (2000).

12. Andreasen, G., Schilardi, P. L., Azzaroni, O., & Salvarezza R. C. Thermal annealing of patterned metal surfaces. *Langmuir* **18,** 10430 (2002).

13. Bedanta, S. & Kleemann, W. Supermagnetism. *J. Phys. D: Appl. Phys.* **42,** 013001 (2009).

14. Chen, Xi *et al.* Relaxation and aging of a superferromagnetic domain state. *Phys. Rev. B*

**68**, 054433 (2003)**.**

15. Skuryev, V. *et al.* Beating the superparamagnetic limit with exchange bias. *Nature* **423,** 850-853 (2003).

16. Chen, G. *et al.* Four-fold magnetic anisotropy induced by the antiferromagnetic order in FeMn/Co/Cu (001) system. *J. Appl. Phys.* **108,** 073905 (2010).

17. Izhikevich, E. M. *Dynamical Systems in Neuroscience: The Geometry of Excitability and Bursting* (MIT Press, Massachusetts, 2007).

18. Sen, P., Ghosh, J., Abdullah, A., Kumar, P., & Vandana. Preparation of Cu, Ag, Fe and Al nanoparticles by the exploding wire technique. *Proc. Indian. Acad. Sci. (Chem. Sci.)* **115,** 499-508 (2003).

**Acknowledgements**

The authors thank the Advanced Instrumentation and Research Facility, Jawaharlal Nehru University, New Delhi for help during the TEM and XRD measurements. We thank Prof. R. Chatterjee of I. I. T., New Delhi and Dr. V. P. S. Awana of N.P.L., New Delhi for allowing the use of SQUID and PPMS facilities, respectively. S.P. Pal thanks CSIR, India, for a research fellowship.

**Author contributions**

P. Sen supervised all aspects of the research. S. P. Pal and P. Sen synthesized and characterized the samples. S. P. Pal analysed the data with inputs from P. Sen. P. Sen wrote the manuscript with comments from S. P. Pal. The manuscript reflects the contributions of both authors.

**Supplementary Information 1. Infusion of crystalline order by annealing of Fe nanoparticles in the presence of activated carbon**

All the X-ray powder diffraction patterns were recorded using a PANalytical X'pert PRO Difffractometer using CuK$_\alpha$ radiation ($\lambda$ =1.5406 A$^0$), and operated at 40 kV and 30 mA filament current. Diffraction scans were recorded with a measurements step of 0.0080 in the range of 2θ=20-80$^0$. Following anneal of a composite of Fe-nano with activated carbon, vacancies are created in the amorphous capping layer through CO$_x$ formation [S1], while thermal excitation (at elevated temperatures) can initiate vacancy migration and lattice rearrangement in the bulk of the nanoparticles. In Fig. S1, evolution of the XRD peaks for Fe-nano-ox annealed at various temperatures, starting from room temperature (300 K) upto 900 K, is shown. This XRD pattern clearly shows the transformation from disrupted lattice to a crystalline oxidized phase, after annealing above 600K. The peaks at 2θ = 30.20$^0$, 35.60$^0$, 43.19$^0$, 57.11$^0$, and 62.67$^0$ correspond to reflections from (220), (311), (400), (511) and (440) planes respectively, of Fe$_3$O$_4$ or γ-Fe$_2$O$_3$ with a face centered cubic lattice. The XRD pattern cannot distinguish between Fe$_3$O$_4$ and γ-Fe$_2$O$_3$ as both occur in the spinel form with comparable d-values. However, the occurrence of these is verified by employing $^{57}$Fe Mössbauer measurements. As we know, two sextets exist in the Mössbauer spectrum of Fe$_3$O$_4$, one corresponding to the tetrahedral site Fe$^{3+}$ ion and the other to an average oxidation state of Fe$^{2.5+}$ ion in the octahedral site. But, Mössbauer spectrum of Fe-nano-ox recorded by us gives only one sextet; the absence of two sextets and the Mössbauer parameters obtained from the spectrum, with almost negligible value of quadrupole shift, affirms presence of the γ-Fe$_2$O$_3$ form. No Fe-C compound formation is observed.

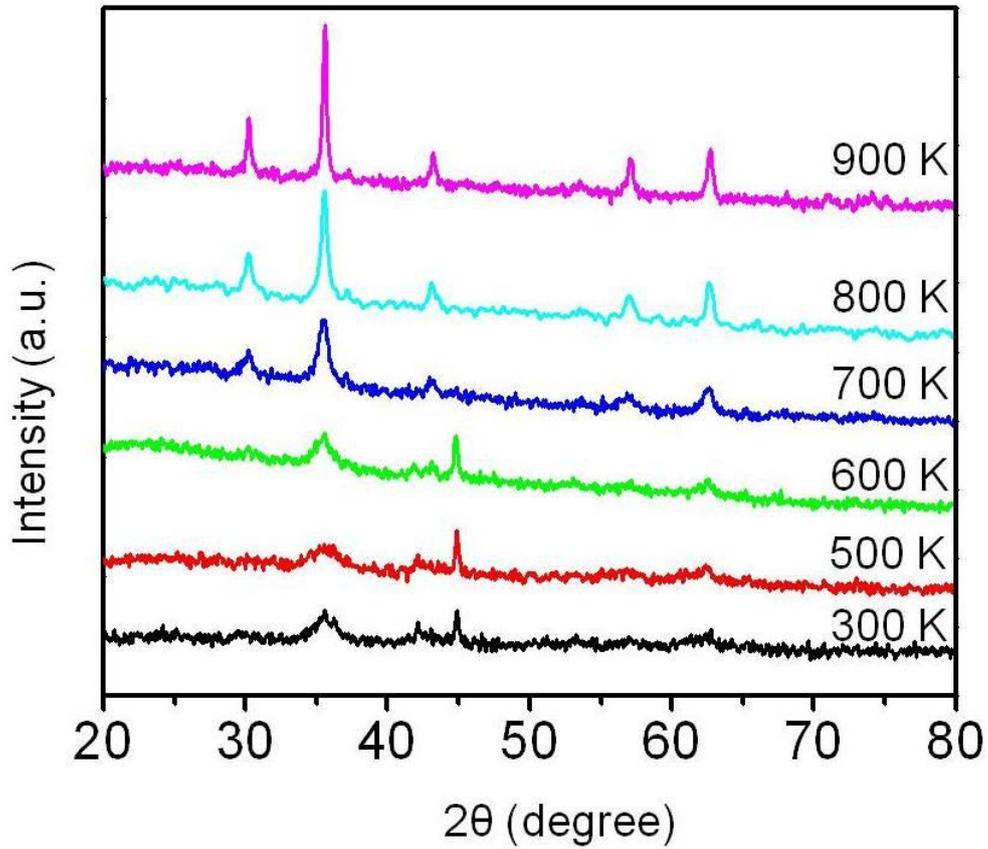

**Figure S1**| XRD pattern of Fe-nano-ox annealed at various temperatures, starting from room temperature (300 K) to 900 K.

**Supplementary Information 2: Magnetization dynamics**

Fig. S2 shows the room temperature M-H curve. This was measured using a physical property measurement system (PPMS-14T, Quantum Design), recorded at 300 K with a sweep rate of 60 Oe min$^{-1}$ which is larger than the sweep rates reported in the main text. The reported magnetization data in the text were collected using a commercial Quantum Design SQUID magnetometer (MPMS-XL) with a sensitivity of $10^{-8}$ emu. A "Reciprocating sample transport option" (RSO) was used in these measurements to achieve better sensitivity. However, the sweep rates available in SQUID were not sufficient to observe the large hysteresis recorded here with the PPMS. Since the magnetic spins have their own dynamics which is slow as compared to the field sweep, they lag behind producing a coercivity value of the 139.0 Oe. Inset shows the full hysteresis loop, swept through ±20 kOe, up to saturation magnetisation.

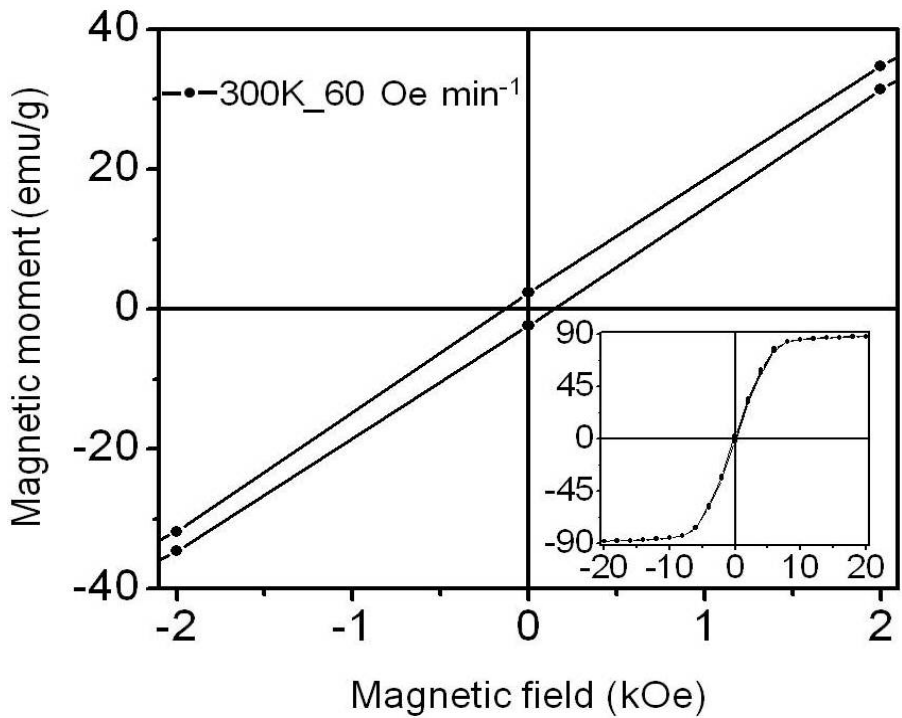

**Figure S2|** Hysteresis loop for Fe-nano taken at 300 K at a sweep rate of 60 Oe min$^{-1}$.

**REFERENCE**

[S1] Castro, C. S., Guerreiro, M. C., Goncalves, M., Oliveira, L. C.A., Anastacio, A. S. Activated carbon/iron oxide composites for the removal of atrazine from aqueous medium. *J. Hazard. Mater.* **164,** 609-614 (2009).